\documentclass[
    aps,
    prl,
    twoside,
    twocolumn,
    10pt,
    floatfix,
    showpacs,
    citeautoscript,
    superscriptaddress,
]{revtex4-2}

\usepackage{amsmath}
\usepackage{amssymb}
\usepackage{booktabs}
\usepackage{comment}
\usepackage{glossaries}
\usepackage{graphicx}
\usepackage[version=4]{mhchem}
\usepackage[per-mode=reciprocal]{siunitx}
\usepackage{tabularx}
\usepackage[dvipsnames]{xcolor}

\usepackage[colorlinks=true]{hyperref}
\hypersetup{
    pdffitwindow=false,
    pdfstartview={FitH},
    pdfnewwindow=true,
    colorlinks=true,
    linkcolor=NavyBlue,
    citecolor=NavyBlue,
    filecolor=NavyBlue,
    urlcolor=NavyBlue,
}

\hypersetup{pdfauthor={E. Berger, N. Khosravian, F. A. A. Nugroho, J. Löfgren, C. Langhammer, P. Erhart}}
\hypersetup{pdftitle={Revealing the kinetics of interfacial surfactant phase transitions through multiscale simulations and in-situ plasmonic sensing}}

\setacronymstyle{long-short}
\newacronym{ctab}{CTAB}{cetyltrimethylammonium bromide}
\newacronym{dft}{DFT}{density functional theory}
\newacronym{fdtd}{FDTD}{finite-difference time-domain}
\newacronym{lspr}{LSPR}{localized surface plasmon resonance}
\newacronym{md}{MD}{molecular dynamics}
\newacronym{si}{SI}{supporting information}

\DeclareSIUnit\angstrom{\text{Å}}
\DeclareSIUnit\Molar{\text{M}}
\DeclareSIUnit\bar{bar}
\renewcommand{\epsilon}[0]{\varepsilon}

\makeatletter
\let\oldtheequation\theequation
\renewcommand\tagform@[1]{\maketag@@@{\ignorespaces#1\unskip\@@italiccorr}}
\renewcommand\theequation{(\oldtheequation)}
\makeatother

\newcommand{\addphyschalmers}{
    Department of Physics and Astronomy,
    Chalmers University of Technology,
    41296 Gothenburg, Sweden
}
\newcommand{\addphysindonesia}{
    Department of Physics,
    Faculty of Mathematics and Natural Sciences,
    Universitas Indonesia, 
    16424 Depok, Indonesia
}
\newcommand{\addinasmart}{
    Institute for Advanced Sustainable Materials Research and Technologies (INA-SMART),
    Faculty of Mathematics and Natural Sciences,
    Universitas Indonesia, 
    16424 Depok, Indonesia
}

\begin{document}

\title{Revealing the kinetics of interfacial surfactant phase transitions\texorpdfstring{\\}{}through multiscale simulations and \textit{in-situ} plasmonic sensing}

\author{Esmée Berger}
\affiliation{\addphyschalmers}
\author{Narjes Khosravian}
\affiliation{\addphyschalmers}
\author{Ferry Anggoro Ardy Nugroho}
\affiliation{\addphysindonesia}
\affiliation{\addinasmart}
\author{Joakim Löfgren}
\affiliation{\addphyschalmers}
\author{Christoph Langhammer}
\affiliation{\addphyschalmers}
\author{Paul Erhart} \email{erhart@chalmers.se}
\affiliation{\addphyschalmers}

\date{\today}

\begin{abstract}
Surfactant self-assembly at solid--liquid interfaces governs interfacial stability, transport, and reactivity across many technologies, yet resolving interfacial surfactant phases and their transition kinetics \textit{in situ} remains challenging.
Here, we establish an atomistically grounded plasmonic framework that quantitatively maps interfacial surfactant \emph{phases} and \emph{phase transitions} onto optical signatures.
Distinct morphologies differ in packing and hydration, modifying the effective permittivity within the optical near field and producing surfactant phase-specific plasmonic extinction peak shifts.
Using cetyltrimethylammonium bromide on silica as a prototypical surfactant--surface system, we combine atomistic simulations, electronic-structure calculations, and continuum electrodynamics to translate molecular morphologies into predicted spectral shifts for literature-reported surface phases.
We experimentally confirm the predicted ordering and magnitude of steady-state peak shifts during stepwise concentration changes, and extract transition kinetics from exponential relaxations of the time-resolved peak shift.
A key mechanistic signature is reversal of the spectral shift direction upon transition from an impermeable bilayer to a water-accessible, channel-containing phase, consistent with hydration-driven reduction of the local effective permittivity.
Because the approach relies on dielectric contrast in the plasmonic near field and works through a dielectric overlayer, it provides a broadly applicable route for real-time identification of interfacial surfactant phases and their kinetics in aqueous conditions.
\end{abstract}

\maketitle

\section{Introduction}

Surfactants govern the structure and dynamics of solid–liquid and liquid–liquid interfaces across a wide range of technological processes, including nanostructure templating \cite{ZhaElzLi19, PenPenWan23, XieRenJia23, WeiZhaDon23, PenPenLi23, PauCha24, YamKurito24}, pharmaceutical synthesis \cite{WanZhaZha21, BoyGonWel21, CheZheZha23, AbdAhmSad24}, perovskite solar cell stabilization \cite{WanYinZhe23}, and microrobotics \cite{ToyMarHan09, BunRanMaa24}.
Their functionality arises from spontaneous self-assembly at interfaces, where amphiphilic molecules reorganize to minimize interfacial free energy and thereby form distinct surface morphologies \cite{KroHolLin2014SurfChem}.
These morphologies govern mechanical stability, transport, chemical reactivity, and templating behavior, such that predictive control of surfactant-mediated processes requires knowledge of interfacial phase diagrams and the kinetics of transitions between surfactant surface phases.

Currently, experimental access to this information remains limited.
Atomic force microscopy provides high spatial resolution but may perturb soft interfaces and offers restricted temporal resolution \cite{Duc02}.
Ensemble-averaging techniques such as ellipsometry or quartz crystal microbalance measurements yield thickness or mass but cannot unambiguously distinguish morphologies with similar coverage \cite{SchReeSek24, BorHoo10}, while scattering and reflectometry approaches require specialized instrumentation and are not readily suited for continuous real-time measurements under dynamic aqueous conditions \cite{SchReeSek24}.
Plasmonic sensing can in principle be used to track adsorption-driven changes in surfactant layers \cite{MeeCelSch16} but has not yet been used to its potential.
Consequently, resolving interfacial surfactant phases \textit{in situ} and tracking their transitions remains a central challenge.

\begin{figure*}
    \centering
    \includegraphics{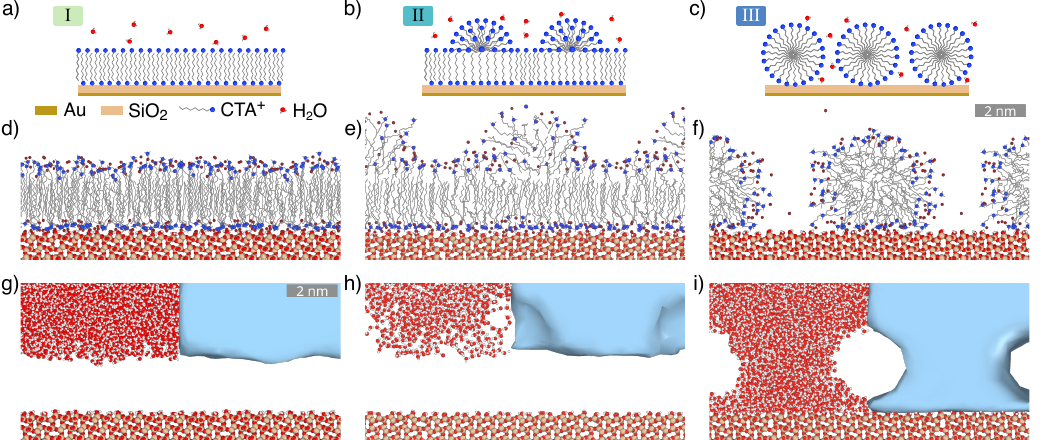}
    \caption{
    \textbf{Atomistic view of CTAB surface phases.}
    (a--c) Schematic representations of CTAB surface phases on silica according to Kadirov \textit{et al.} \cite{KadLitNiz14}, shown in order of increasing concentration: bilayer (I), hemispherically capped bilayer (II), and cylindrical micelles (III).
    (d--f) Corresponding atomistic MD snapshots illustrating the molecular organization of each phase at the solid–liquid interface.
    (g--i) Spatial distribution of water for the same structures.
    Water is shown as individual atoms (left) and as a density isosurface (right), illustrating the pronounced differences in interfacial hydration and water accessibility between phases.
    These variations in local water content give rise to distinct effective dielectric environments within the optical near field.
    The blue isosurfaces were generated using OVITO \cite{Stu10}.
    The length scale is indicated by the scale bars in (f,g).
    }
    \label{fig:atomistic view}
\end{figure*}

Here, we demonstrate a strategy based on plasmonic sensing that probes interfacial surfactant phases via their dielectric fingerprints, which we show can be derived from atomistic simulations.
Differences in packing, porosity, and hydration between morphologies determine the effective permittivity within the nanometer-scale region adjacent to the surface.
Phase transitions therefore induce characteristic changes in the local dielectric environment that can be detected as shifts in the extinction spectrum of plasmonic nanostructures.
This near-field readout enables continuous, non-contact measurements under fully aqueous conditions and provides direct access to both equilibrium surface phases and their transformation kinetics.

We validate this concept using \gls{ctab} adsorbed on silica as a prototypical surfactant–surface system.
\Gls{ctab} is one of the most extensively studied cationic surfactants and plays a central role in nanoparticle synthesis and colloidal stabilization \cite{NikEl-03, LiuGuy05, VerPor24}.
Although its bulk phase diagram is well established \cite{CopGiaNic04, KriGhoLak05}, interfacial phase behavior remains less well understood due to the difficulty of probing buried solid–liquid interfaces.
Using atomic force microscopy, Kadirov \textit{et al.} reported a concentration-dependent transition from an impermeable bilayer to a permeable cylindrical surface phase on silica \cite{KadLitNiz14}.
This structurally non-trivial transition entails a qualitative change in interfacial hydration, from water exclusion to water-accessible domains, and is therefore expected to produce distinct dielectric signatures within the optical near field.

By combining multiscale modeling with experiment, we quantitatively link interfacial morphology to plasmonic response and identify the microscopic origin of the observed spectral shifts.
The agreement between predicted and measured signatures during controlled variation of surfactant concentration establishes a framework in which interfacial phase transitions are mapped onto measurable dielectric contrasts.

Because the method relies solely on changes in near-field permittivity, it is independent of specific molecular chemistry.
Although demonstrated for \gls{ctab} on silica, the approach can be expected to be broadly applicable to diverse surfactant–surface systems and offers a route toward real-time mapping of interfacial phase diagrams and transition kinetics in soft and nanoscale materials.

\begin{figure*}
    \centering
    \includegraphics{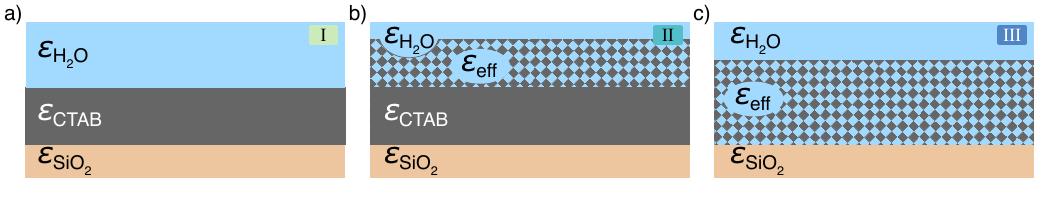}
    \caption{
    \textbf{Continuum representations of the CTAB surface phases used in the FDTD simulations.}
    (a) Bilayer phase (I), (b) hemispherically capped bilayer phase (II), and (c) cylindrical micelle phase (III).
    The interfacial region is represented as layers parallel to the surface.
    Pure \gls{ctab} layers ($f=0$) and bulk water ($f=1$) are assigned their respective permittivities, while mixed layers are described by an effective permittivity $\epsilon_\text{eff}$ obtained from the layer-resolved water volume fraction $f$ according to \autoref{eq:effective-permittivity}.
    In phase I, the layer adjacent to silica consists of pure CTAB; in phase II, an additional mixed layer represents the caps; and in phase III, a mixed layer is present directly at the surface due to water-accessible channels.
    }
        \label{fig:continuum}
\end{figure*}

\section{Results and Discussion}

For \gls{ctab} adsorbed on silica---the prototypical surfactant–solid interface under consideration---a surface phase diagram has been proposed based on atomic force microscopy measurements \cite{KadLitNiz14}.
Upon increasing the \gls{ctab} concentration from \qty{0}{\milli\Molar} to above \qty{1}{\milli\Molar}, qualitatively distinct surface morphologies are expected to form.
Between \qtyrange{0}{0.5}{\milli\Molar}, an impermeable bilayer phase (I; \autoref{fig:atomistic view}a) has been reported.
In the range \qtyrange{0.5}{0.7}{\milli\Molar}, this bilayer transforms into a hemispherically capped bilayer phase (II; \autoref{fig:atomistic view}b), and for concentrations exceeding \qty{1}{\milli\Molar}, a permeable cylindrical phase containing water channels is formed (III; \autoref{fig:atomistic view}c).
These phases differ not only in thickness and surfactant density, but also in the spatial distribution of water within the interfacial layer.
Such structural differences are expected to modify the effective permittivity in the near-surface region.
Given that the \gls{lspr} response is highly sensitive to changes in the local dielectric environment, these phase transitions are anticipated to produce measurable peak shifts in the extinction spectrum \cite{Nug22, EkbRahRos22}.

To quantify this expectation, we employ a multiscale modeling framework that connects electronic structure calculations, atomistic simulations, and continuum electrodynamics.
At the atomic scale, the relevant \gls{ctab} surface phases are modeled using \gls{md} simulations.
The resulting equilibrium morphologies then provide geometric parameters and water distributions that are subsequently translated into continuum representations used in \gls{fdtd} simulations.
Combining these geometries with permittivities obtained from electronic structure calculations enables computation of extinction spectra, which constitute the experimentally accessible observable in plasmonic sensing measurements.
Finally, we confirm these predictions through targeted experiments, and exploit this sensing approach to quantify the kinetics of the transitions between different phases.

\subsection{Atomistic Simulations}

Atomistic \gls{md} simulations were performed to resolve the morphology and hydration characteristics of the proposed \gls{ctab} surface phases.
Because the length and time scales accessible in these simulations preclude direct simulation of macroscopic adsorption equilibria, we initialize systems with varying surface coverages to obtain the relevant interfacial structures.
Equilibration of these systems yields stable morphologies corresponding to phases I--III (\autoref{fig:atomistic view}d--f), which allow us to determine the corresponding spatial distribution of \gls{ctab} and water (\autoref{fig:atomistic view}g--i).
The simulated layer heights and characteristic dimensions are consistent with the experimentally reported values within the substantial uncertainty of the atomic force microscopy measurements.

The simulations reveal pronounced morphological differences between the phases.
In the bilayer (I) and hemispherically capped bilayer (II) phases, the surfactant layer largely excludes water from the region immediately adjacent to the silica surface.
In contrast, the cylindrical phase (III) exhibits water-accessible channels that penetrate the surfactant layer.
These differences in interfacial hydration imply distinct effective dielectric environments within the optical near field.
The atomistic simulations thereby provide the structural basis for predicting phase-dependent variations in the plasmonic response.

\begin{figure*}
    \centering
    \includegraphics{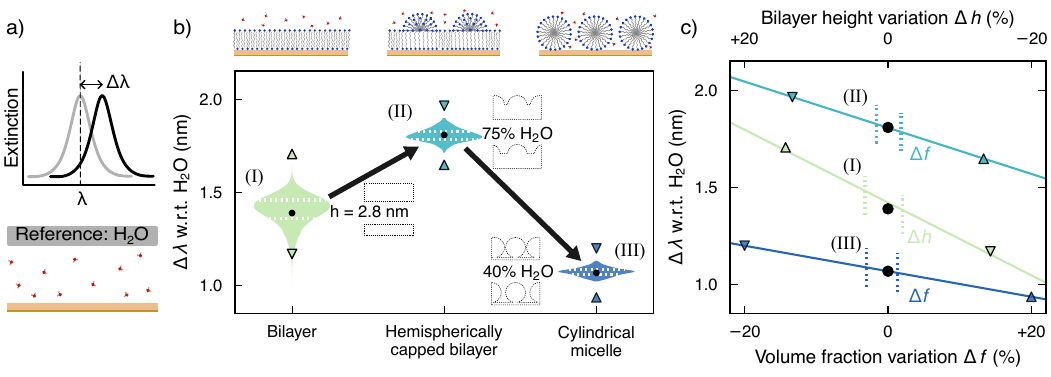}
    \caption{
    \textbf{Simulated plasmonic response and FDTD sensitivity analysis.}
    (a) Schematic extinction spectrum (top) illustrating the sensing principle.
    The peak shift is defined as $\Delta\lambda=\lambda_\text{peak}-\lambda_\text{ref}$, where $\lambda_\text{ref}$ is the extinction peak position of a water-only reference surface (bottom).
    (b) Predicted \gls{lspr} peak shifts $\Delta\lambda$ obtained from the \gls{fdtd} simulations of the continuum geometries for the bilayer (I), hemispherically capped bilayer (II), and cylindrical micelle (III) phases.
    The colored violin plots indicate the probability distributions obtained from \gls{md} simulations of (I) bilayer height $h$, (II) water volume fraction $f$ in the mixed cap layer, and (III) water volume fraction $f$ in the cylindrical phase.
    The dashed lines indicate the \qty{25}{\percent} and \qty{75}{\percent} quantiles. 
    The black dots mark the peak shifts for the respective most likely configurations, while the triangles indicate the range of values tested in the \gls{fdtd} simulations.
    (c) Parameter sweeps performed in the \gls{fdtd} simulations to assess the sensitivity of $\Delta\lambda$ to (I) bilayer height $h$, (II) water volume fraction $f$ in the mixed cap layer, and (III) water volume fraction $f$ in the cylindrical phase.
    The circles and triangles refer to the respective data points in (b) and the dashed lines correspond to the \qty{25}{\percent} and \qty{75}{\percent} quantiles of the probabiliy distributions in (b).
    Increasing the layer height increases $\Delta\lambda$, whereas increasing water fraction decreases $\Delta\lambda$.
    The opposing trends for thickness and hydration explain the sign change in peak shift between phases II and III.
    }
    \label{fig:sensitivity-scan}
\end{figure*}

\subsection{Simulated Dielectric Response}

To compute extinction spectra for the different interfacial morphologies, we performed \gls{fdtd} simulations using continuum representations of the surfactant surface phases derived from the atomistic structures.
The resulting geometries are parameterized by the surfactant-layer height and the spatial distribution of water within the interfacial region (\autoref{fig:continuum}), as obtained from the \gls{md} simulations.

The continuum models describe the interfacial region as a stack of layers parallel to the surface, each assigned an effective permittivity.
In layers containing both \gls{ctab} and water, we approximate the local dielectric response using an effective-medium description,
\begin{equation}\label{eq:effective-permittivity}
    \epsilon_\text{eff}=f\epsilon_\text{H$_2$O} + (1-f)\epsilon_\text{CTAB},
\end{equation}
where $f$ is the layer-resolved volume fraction of water (i.e., the fraction of the layer volume occupied by water, as determined from the atomistic structure).
This definition allows the continuum models to capture the key physical distinction between impermeable and permeable phases in a minimal way: in the bilayer phase (I), the layer adjacent to the silica surface is purely \gls{ctab} ($f=0$) with bulk water above ($f=1$), whereas the capped bilayer (II) contains an intermediate mixed layer ($0<f<1$) representing the caps, and the cylindrical micelle phase (III) contains a mixed layer already at the surface due to water-accessible channels (\autoref{fig:continuum}).

The use of an effective-medium approximation is justified here by the separation of length scales relevant to the optical response.
The characteristic feature sizes within the surfactant layer (e.g., channel/cap dimensions) are much smaller than the optical wavelength and smaller than the near-field probing depth of the plasmonic sensor, such that the measured response is dominated by an averaged dielectric environment over the interfacial region.
To ensure that the conclusions do not depend on the specific mixing rule, we compare alternative effective-medium schemes (e.g., Maxwell--Garnett and Bruggeman) in the SI, finding that the qualitative trends reported below, including the change in peak-shift direction for the impermeable-to-permeable transition, are robust (\autoref{sfig:mixing-rules}).

The permittivity of water, $\epsilon_\text{H$_2$O}$, is taken from tabulated, wavelength-dependent optical data.
Determining $\epsilon_\text{CTAB}$ requires additional attention because the dielectric response depends on the molecular packing density of the surfactant phase.
We therefore compute the frequency-dependent dielectric function of bulk \gls{ctab} using \gls{dft} for a range of representative bulk densities. 
Over the density range relevant to the simulated surface phases, the real part of $\epsilon_\text{CTAB}$ varies only weakly with density, but the trend is systematic and is included in the continuum modeling (\autoref{sfig:ctab-permittivity}).
Previous experimental studies report $\epsilon_\text{CTAB}\approx2$ for \gls{ctab} without accounting for this density dependence \cite{YuVarIru07a, MeeCelSch16}.
Using the phase-resolved densities extracted from the atomistic structures (\autoref{fig:atomistic view}d--f), we adopt $\epsilon_\text{CTAB}\approx2.6$ for phases I--II and $\epsilon_\text{CTAB}\approx2.4$ for phase III in the \gls{fdtd} simulations, consistent with \autoref{sfig:ctab-permittivity}.
We note that both water and \gls{ctab} are wide-gap materials, and accordingly the real part of their dielectric functions is smooth across the spectral range of interest, without sharp structure that would qualitatively alter the trends discussed here.

The simulated extinction spectra are analyzed in terms of the \gls{lspr} peak shift, $\Delta\lambda$, relative to the spectrum in the absence of \gls{ctab} (\autoref{fig:sensitivity-scan}a).
The resulting phase-dependent peak shifts (black dots in \autoref{fig:sensitivity-scan}b) are distinct, even when accounting for the uncertainty in the geometric parameters extracted from the atomistic simulations (violin plots in \autoref{fig:sensitivity-scan}b).
Specifically, adsorption of the impermeable bilayer (I) produces a positive peak shift relative to the water-only reference, the peak shift increases further for the capped bilayer (II), and the peak shift decreases for the cylindrical micelle phase (III).
This establishes that plasmonic sensing can, in principle, distinguish between the proposed \gls{ctab} surface phases based on their dielectric signatures.

A key result is that the \emph{direction} of the peak shift changes between phases II and III.
Although the \gls{ctab} concentration increases from phase I to II and from phase II to III (\autoref{fig:atomistic view}a--c), the simulated optical response differs qualitatively in the latter case.
This behavior indicates that the II$\rightarrow$III transition is not simply a thickening or densification of the interfacial layer, but is instead dominated by the emergence of water-accessible domains that reduce the effective permittivity sampled by the plasmonic near field.
To disentangle the geometric and compositional contributions to $\Delta\lambda$ and to rationalize the sign change in the response, we next perform a sensitivity analysis in which we vary (i) the bilayer height $h$ and (ii) the water volume fraction $f$ within the mixed interfacial layer for phases II and III (\autoref{fig:sensitivity-scan}c).
This is a sensible approach since these variations solely affect the interfacial hydration level without altering the respective geometries of each phase.

\begin{figure*}
    \centering
    \includegraphics{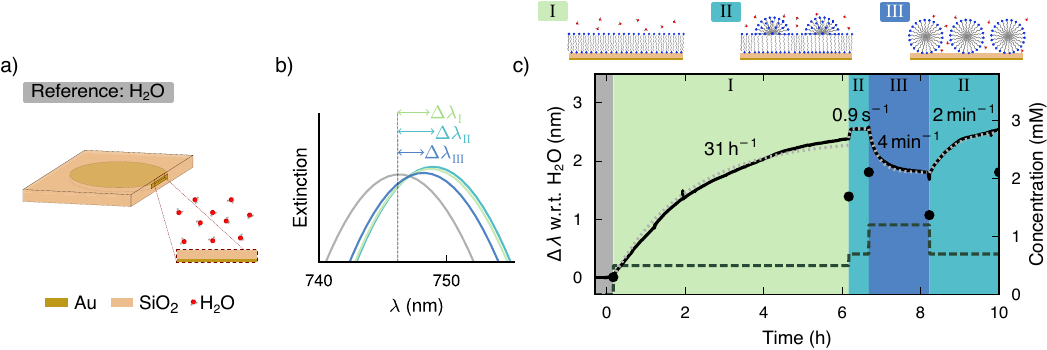}
    \caption{
    \textbf{Operating principle and experimental plasmonic response to interfacial CTAB-on-silica phases, with extracted apparent rate constants.}
    (a) Schematic view of the \acrfull{lspr} sensor. (b) Representative extinction spectra, illustrating how changes in the interfacial dielectric environment shift the plasmonic peak.
    (c) Measured $\Delta\lambda(t)$ (solid black line) during stepwise changes of \gls{ctab} concentration (dashed green line), showing relaxation to a plateau after each step.
    These plateaus are in quantitative agreement with the simulated steady-state peak shifts (black dots) from \autoref{fig:sensitivity-scan}.
    Notably, increasing the concentration from \qty{0.7}{\milli\Molar} to \qty{1.2}{\milli\Molar} decreases $\Delta\lambda$, consistent with a transition from an impermeable phase (II) to a permeable, channel-containing phase (III).
    Apparent rate constants $k_\text{app}=1/\tau$ for the CTAB surface-phase transitions are obtained from exponential relaxations of the \gls{lspr} response (dotted gray lines mark single-exponential fits). 
    The transition from the water-only reference surface to the bilayer phase (I) is the slowest, whereas the I$\rightarrow$II transition (bilayer to capped bilayer) is the fastest.
    }
    \label{fig:schematic+Kadirov+experiment}
\end{figure*}

The bilayer height scan (\autoref{fig:sensitivity-scan}c, I) shows that increasing the surfactant-layer thickness increases the peak shift.
This follows from the increased overlap of the plasmonic near field with the higher-permittivity surfactant region and the corresponding displacement of water away from the surface.
Conversely, increasing the water fraction $f$ within the interfacial layer reduces the peak shift, as observed for both the capped bilayer (\autoref{fig:sensitivity-scan}c, II) and the cylindrical micelle phase (\autoref{fig:sensitivity-scan}c, III).
These trends explain the qualitative difference between the I$\rightarrow$II and II$\rightarrow$III transitions.
In the absence of channel formation, an increased surfactant loading primarily thickens the layer and thus increases $\Delta\lambda$, consistent with the I$\rightarrow$II transition.
By contrast, a decrease in $\Delta\lambda$ upon increasing concentration, as observed for II$\rightarrow$III, requires a reduction of the local effective permittivity, which is naturally achieved by introducing water-accessible domains within the interfacial structure.
Importantly, the qualitative sign change for the impermeable-to-permeable transition is robust to the precise value of $\epsilon_\text{CTAB}$ and persists across the explored parameter range (\autoref{sfig:fdtd-ctab-permittivity-scan}), supporting its generality as an optical signature of hydration-driven morphological reorganization rather than a consequence of a particular dielectric parameter choice.

\subsection{Experimental Validation by LSPR Measurements}

Motivated by the multiscale modeling, we experimentally validate that plasmonic sensing is sensitive to the interfacial adsorption/rehydration state of \gls{ctab}, that the steady-state peak shifts follow the predicted phase-dependent ordering, and that the response is reversible.
We monitor the \gls{lspr} peak shift, $\Delta\lambda$, relative to its initial value (\autoref{fig:schematic+Kadirov+experiment}a) while varying the \gls{ctab} concentration in a stepwise manner (\autoref{fig:schematic+Kadirov+experiment}c).
Each concentration step produces a clear spectral response (\autoref{fig:schematic+Kadirov+experiment}b, c), and $\Delta\lambda$ relaxes on timescales ranging from minutes to hours until reaching a plateau (\autoref{fig:schematic+Kadirov+experiment}c).

While the transition kinetics are analyzed in detail below, we first compare the plateau values at the end of each concentration interval to the \gls{fdtd}-predicted peak shifts extracted from the simulated extinction spectra (black dots in \autoref{fig:schematic+Kadirov+experiment}c).
The experimentally measured plateau shifts show qualitative agreement with the simulated ordering across the concentration regimes, demonstrating sensitivity to phase-dependent changes in interfacial morphology and hydration.
In addition, the experimental plateaus match the predicted \gls{lspr} peak shift magnitudes quantitatively.
That both the ordering and the magnitude of the predicted steady-state peak shifts are confirmed by the experiment strongly corroborates our theoretical approach and explanation of the underlying microscopic mechanism.
Moreover, the saturated response is reversible under the concentration protocol employed here.
Upon decreasing the concentration from \qty{1.2}{\milli\Molar} back to \qty{0.7}{\milli\Molar}, $\Delta\lambda$ returns to the plateau value previously recorded at \qty{0.7}{\milli\Molar} within experimental uncertainty.

\subsection{Kinetics of the Phase Transitions}

A key advantage of plasmonic sensing is that it provides continuous, time-resolved access to the kinetics of adsorption and phase transitions within the interfacial surfactant layer without interfering with the dynamics.
For each stepwise change in \gls{ctab} concentration, we analyze the relaxation of the \gls{lspr} peak shift, $\Delta\lambda(t)$, toward its new plateau value (\autoref{fig:schematic+Kadirov+experiment}c) and extract a characteristic transition time.
In practice, each relaxation segment is well described by a single-exponential approach to the plateau, from which we obtain a time constant $\tau$.
From this characteristic time, we report the apparent rate constant as $k_\mathrm{app}=1/\tau$ for each transition (labels in \autoref{fig:schematic+Kadirov+experiment}c).

The adsorption step from the water-only reference surface to formation of the bilayer phase (I) is the slowest transition, with an apparent rate constant approximately two orders of magnitude smaller than that of the I$\rightarrow$II transition.
The remaining transitions between the capped bilayer (II) and cylindrical micelle phase (III), as well as the reverse transition, exhibit apparent rate constants of the same order of magnitude and lie between these two extremes.

The broad distribution of time scales suggests that the interfacial \gls{ctab} layer can, under certain conditions, become kinetically trapped in metastable morphologies on experimental time scales.
This has practical implications for surfactant-mediated processes, where the functional response may depend not only on the equilibrium surface phase but also on the available equilibration time following changes in concentration or operating conditions.

\section{Conclusions}

We have demonstrated that interfacial surfactant phases can be distinguished through their dielectric fingerprints using plasmonic sensing.
Because different morphologies modify the local permittivity within the optical near field, phase transitions are translated into measurable shifts of the \gls{lspr} resonance.
This provides a non-contact, label-free route to probe surfactant structure \textit{in situ} under fully aqueous conditions.

By combining multiscale modeling with experiment, we established that plasmonic sensing enables monitoring of both the equilibrium \gls{ctab} surface phases on silica and the kinetics of the transitions between them.
The experimentally observed steady-state plasmonic peak shifts follow the ordering predicted by the simulations, and the response is reversible under the applied concentration protocol.
Importantly, the qualitative trends in the simulated peak shifts, including the reversal in the direction of the response for the impermeable-to-permeable transition, are robust with respect to the specific numerical value of the \gls{ctab} permittivity used in the continuum description.
As the approach relies only on dielectric contrast within the near-field region, it is broadly applicable to surfactant–surface systems in which the surfactant and surrounding medium exhibit distinct permittivities.

The ability to directly observe transitions between impermeable and water-accessible surfactant morphologies \textit{in situ} has practical implications for systems in which interfacial structure governs functionality, such as surfactant-mediated nanoparticle growth or heterogeneous catalysis.
Moreover, the wide range of transition time scales observed here highlights the potential for kinetic trapping of metastable surface phases on experimentally relevant time scales.
Real-time plasmonic monitoring therefore offers a means to verify whether a desired interfacial state has been reached before subsequent processing steps are undertaken.

\section{Methodology}

\subsection{Atomistic Simulations}
All \gls{md} simulations were performed using the \textsc{gromacs} package (version 2022.2) \cite{BerVanVan95, BauHesLin22} with the \textsc{gromos96 54a7} force field \cite{SchEicCho11}.
The model systems comprise a silica slab described by the Q3 model parameters from Emami \textit{et al.} \cite{EmaPudBer14a}, \ce{CTA^+} with force-field parameters from the Automated Topology Builder (ATB; entry 366763) \cite{MalZuoBre11, StrCarVis18}, and \ce{Br^-} with parameters from da Silva and Meneghetti \cite{daMen18}.
Initially, a silica slab of dimensions \qtyproduct{6.7 x 7.0 x 2.6}{\nano\meter\cubed} was placed in a simulation cell of height \qty{25}{\nano\meter}, with a \gls{ctab} bilayer on either side of the slab, following previous simulations of related systems \cite{MeeSul16, daMen18}.
The interdigitation of the alkyl tails was \qty{1.8}{\nano\meter}.
To generate different initial configurations, the surface density of \gls{ctab} was varied in the range \qtyrange[per-mode=reciprocal]{1.05}{3.10}{\per\nano\meter\squared}.
The surfactant-covered surfaces were solvated with SPC/E water \cite{BerGriStr87}, and any water molecules initially located within the hydrophobic region of the bilayer were removed prior to equilibration.

Periodic boundary conditions were applied in all directions.
The simulation protocol consisted of steepest-descent energy minimization until the maximum force was below \qty{1000}{\kilo\joule\per\mol}, followed by \qty{500}{\pico\second} of equilibration in the canonical (NVT) ensemble using a velocity-rescaling thermostat \cite{BusDonPar07} at \qty{300}{\kelvin}.
Subsequently, \qty{1000}{\pico\second} of equilibration was performed in the isothermal--isobaric (NPT) ensemble at \qty{300}{\kelvin} and \qty{1}{\bar} using a semi-isotropic Berendsen barostat \cite{BerPosVan84}.
Production simulations were then carried out in the NVT ensemble with a time step of \qty{2}{\femto\second}.
Production run lengths ranged from \qtyrange{250}{400}{\nano\second}, depending on the time required for the interfacial morphology to reach a stable surface phase, and the final \qty{30}{\nano\second} of each trajectory were used for analysis.
Configurations were visualized using \textsc{ovito} \cite{Stu10}.

\subsection{Density Functional Theory Calculations}
The \gls{dft} calculations of bulk \gls{ctab} (8 molecules with periodic boundary conditions) were performed using the projector augmented-wave method \cite{Blo94, KreJou99} with a plane-wave energy cutoff of \qty{520}{\electronvolt}, as implemented in version 6.4.2 of the Vienna \textit{Ab initio} Simulation Package \cite{KreHaf93, KreFur96, KreFur96a}.
The PBE exchange-correlation functional \cite{PerBurErn96} was employed.
An automatically generated $\Gamma$-centered Monkhorst--Pack \textbf{k}-point mesh \cite{MonPac76} with a maximum spacing of \qty[per-mode=reciprocal]{0.2}{\per\angstrom} was used to sample the Brillouin zone.
Frequency-dependent dielectric functions were computed within the independent-particle approximation, with the number of empty bands increased stepwise to a final value of \num{2600} to ensure convergence (\autoref{sfig:ctab-permittivity-nbands-convergence}).

\subsection{Finite-Difference Time-Domain Simulations}
Extinction spectra were computed using \gls{fdtd} simulations carried out with \textsc{meep} (version 1.12.0) \cite{OskRouIba10}.
A cylindrical gold nanodisk with diameter \qty{100}{\nano\meter} and height \qty{20}{\nano\meter} was positioned \qty{20}{\nano\meter} below the top surface of a \qty{140}{\nano\meter} thick silica substrate.
The silica was modeled with a constant refractive index of \num{1.47813}, while gold was described using the dielectric function from Ref.~\citenum{RakDjuEla98} as provided in the materials library in \textsc{meep}.

To represent the surfactant-covered cases, a layer was placed on top of the silica substrate with a height determined from the \gls{md} simulations and an effective permittivity $\epsilon_\text{eff}$ according to \autoref{eq:effective-permittivity}.
Here, $f$ denotes the layer-resolved volume fraction of water, and $\epsilon_\text{CTAB}$ was determined from the \gls{dft} calculations based on the phase-resolved densities extracted from the \gls{md} simulations.
A \qty{100}{\nano\meter} thick water layer was placed above the substrate. 
The small dispersion in the water permittivity over the \qtyrange{1}{5}{\electronvolt} spectral range considered here was neglected and the permittivity $\epsilon_\text{H$_2$O} = 1.3454^2$ at \qty{330}{\nano\meter} was used as a representative value \cite{HalQue73}.
The system was excited with a normally incident Gaussian source spanning \qtyrange{1}{5}{\electronvolt}, which includes the experimentally accessible range.
Perfectly matched layers were applied in all directions, with a minimum distance of \qty{100}{\nano\meter} from the nanodisk.
A spatial resolution of \qty{1}{pixel/\nano\meter} was used, ensuring sufficient accuracy to resolve peak shifts while maintaining computational feasibility.
Peak positions were extracted from the computed extinction spectra using a polynomial fit in a narrow window around the resonance maximum.

\subsection{Localized Surface Plasmon Resonance Measurement}
The \gls{lspr} measurements were conducted using a \qtyproduct{1 x 1}{\centi\meter\squared} sensor chip comprising a glass substrate (Borofloat, Schott Scandinavia) decorated with a quasi-random array of gold nanodisks (\qty{100}{\nano\meter} diameter, \qty{20}{\nano\meter} height) fabricated by hole-mask colloidal lithography \cite{Nug16, Nug17, Nug19}.
Prior to the experiments, the sensor chip was coated with a \qty{20}{\nano\meter} conformal silica layer by chemical vapor deposition (STS PE-CVD).
\Gls{ctab} (Sigma Aldrich, purity $\geq\qty{98}{\percent}$) was diluted in Milli-Q water (Millipore) and introduced at different concentrations using a commercial titanium flow cell with optical access (XNano, Insplorion AB).
Measurements were performed at room temperature under a constant flow of \qty{100}{\micro\liter\per\minute}, regulated by a peristaltic pump (Ismatec).
The sensor chip was illuminated using a fiber-coupled halogen lamp (AvaLight-Hal, Avantes), and extinction spectra were recorded with a fiber-coupled fixed-grating spectrometer (AvaSpec-HS-TEC, Avantes).
Peak positions were obtained by fitting the resonance with a Lorentzian line shape; no baseline correction was required, as only the peak position was extracted.

\section{Acknowledgments}
This work was funded by the Swedish Research Council (grant numbers 2020-04935, 2021-05072).
C.L. has received funding from the European Research Council (ERC) under the European Union’s Horizon Europe research and innovation program (101043480/NACAREI). 
F.A.A.N. is funded by the Indonesian Endowment Fund for Education (LPDP) on behalf of the Indonesian Ministry of Higher Education, Science and Technology and managed under the EQUITY Program (Contract No. 4302/B3/DT.03.08/2025 and 573/PKS/R/UI/2025)
Computer time allocations by the National Academic Infrastructure for Supercomputing in Sweden (NAISS) at NSC, PDC, and C3SE are gratefully acknowledged.
Part of this work was carried out at the Chalmers MC2 my-Fab cleanroom facility.

\end{document}